\documentstyle[aps]{revtex}
\def\be{\begin{equation}}
\def\ee{\end{equation}}
\def\bea{\begin{eqnarray}}
\def\eea{\end{eqnarray}}
\newcommand{\labell}[1]{\label{#1}}
\newcommand{\reef}[1]{(\ref{#1})}
\newcommand{\cO}{{\cal O}}
\newcommand{\cR}{{\cal R}}

\def\IR{{\hbox{{\rm I}\kern-.2em\hbox{\rm R}}}}
\def\IB{{\hbox{{\rm I}\kern-.2em\hbox{\rm B}}}}
\def\IN{{\hbox{{\rm I}\kern-.2em\hbox{\rm N}}}}
\def\IC{\,\,{\hbox{{\rm I}\kern-.59em\hbox{\bf C}}}}
\def\IZ{{\hbox{{\rm Z}\kern-.4em\hbox{\rm Z}}}}
\def\IP{{\hbox{{\rm I}\kern-.2em\hbox{\rm P}}}}
\def\IH{{\hbox{{\rm I}\kern-.4em\hbox{\rm H}}}}
\def\ID{{\hbox{{\rm I}\kern-.2em\hbox{\rm D}}}}
\def\II{{\hbox{\rm I}\kern-.2em\hbox{\rm I}}}

\begin{document}
\title{\begin{flushright}
{\small hep-th/9903238}
\end{flushright}
Surface Terms as Counterterms in the AdS/CFT Correspondence}

\author{Roberto Emparan$^a$, Clifford
V. Johnson$^b$,  Robert C. Myers$^c$} \author{} \address{$^a$
Department of Mathematical Sciences, University of Durham, DH1 3LE,
UK.\\ Departamento de F{\'\i}sica Te\'orica, Universidad del Pa{\'\i}s
Vasco, Apdo. 644, E-48080 Bilbao, Spain.}  \address{$^b$ Department of
Physics and Astronomy, University of Kentucky, Lexington, KY
40506--0055, USA.}  \address{$^c$Physics Department, McGill
University, Montr\'{e}al, PQ, H3A 2T8, Canada.}  \address{\small
$^a$roberto.emparan@durham.ac.uk,
$^b$cvj@pa.uky.edu, $^c$rcm@hep.physics.mcgill.ca}

\author{}

\maketitle \begin{abstract} We examine the recently proposed technique
of adding boundary counterterms to the gravitational action for
spacetimes which are locally asymptotic to anti--de Sitter. In
particular, we explicitly identify higher order counterterms, which
allow us to consider spacetimes of dimensions~$d\le7$. As the
counterterms eliminate the need of ``background subtraction" in
calculating the action, we apply this technique to study examples
where the appropriate background was ambiguous or unknown: topological
black holes, Taub--NUT--AdS and Taub--Bolt--AdS.  We also identify
certain cases where the covariant counterterms fail to render the
action finite, and we comment on the dual field theory interpretation
of this result. In some examples, the case of vanishing cosmological
constant may be recovered in a limit, which allows us to check results
and resolve ambiguities in certain asymptotically flat spacetime
computations in the literature.
\end{abstract}

\section{Introduction}

The AdS/CFT correspondence asserts that there is an equivalence
between a gravitational theory in $d$-dimensional anti--de Sitter
(AdS) spacetime and a conformal field theory (CFT) living in a
$(d{-}1)$--dimensional ``boundary" spacetime\cite{edads}. This
equivalence or duality is best understood in the context of string
theory with $d{=}5$, where the duality relates type~IIB superstring
theory on AdS$_5{\times}S^5$, and ${\cal N}{=}4$ supersymmetric
Yang--Mills theory with gauge group $SU(N)$ in four
dimensions~\cite{juan,gubkleb}.  The precise formulation of the
AdS/CFT correspondence is made by equating the partition functions of
the two theories:
\be
Z_{AdS}(\phi_{0,i})=Z_{CFT}(\phi_{0,i})\ .
\labell{zz}
\ee
Here the fields $\phi_{0,i}$ have two interpretations: On the gravity
side, these fields correspond to the boundary data or boundary values
(up to a certain rescaling \cite{edads}) for the bulk fields $\phi_i$
which propagate in the AdS space. On the field theory side, these
fields correspond to external source currents coupled to various CFT
operators. Thus correlation functions of the operators in the CFT can
be determined through a calculation using the dynamics of gravity in
AdS spacetime \cite{edads,gubkleb}.  In certain instances, one can
consider evaluating the AdS partition function in a saddle--point
approximation:
\be
e^{-I_{AdS}(\phi_i)}=\left\langle e^{\int \phi_{0,i}\cO^i}
\right\rangle_{CFT}
\labell{genie}
\ee
where $I_{AdS}(\phi_i)$ is the classical gravitational action as a
functional
of the (super)gravity fields, and $\cO^i$ are the dual CFT operators.
Hence in this approximation, the AdS action becomes the generating
function of the connected correlation functions in the CFT
\cite{edads,gubkleb}. This framework is also naturally extended
to considering CFT states for which certain operators acquire
expectation values by considering solutions of the gravitational
equations which are only asymptotically AdS \cite{edadsii,alba}.

One aspect of the duality which will be interesting for the present
investigation is the choice of the background metric $\gamma_{ij}$
required to define the field theory. This metric is related by an
infinite conformal transformation to the induced metric $h_{ij}$ on
the boundary of the AdS spacetime \cite{edads}. Since the boundary
conformal transformation is divergent, one regularizes the calculation
by considering the induced metric for a family of surfaces which
approach the boundary in a limit.  This regularization procedure then
will depend on the choices of coordinates in the asymptotic AdS
region, {\it i.e.,} it depends on the precise family of surfaces
chosen. With different choices, the background geometry inherited
by the CFT takes a completely different form. For example, depending
on the choice of radial slicing for AdS$_{n+1}$, the boundary geometry
can be $S^1{\times}S^n$, $S^{n+1}$, $S^1{\times}{\IR}^n$. We will
discuss these and other possibilities in section \ref{sec:faces}. This
procedure therefore allows one to study the CFT with different
background
geometries.  From the point of view of the gravity theory, this
procedure is interesting because naively the expressions on the
left-hand side of eqns.~\reef{zz} and \reef{genie} are coordinate
invariant. However, the asymptotic regularization explicitly breaks
this covariance.

Returning to eqn.~\reef{genie},
considering the gravitational path integral in the saddle-point
approximation
has a long history in the quantum gravity literature, in particular
in context of black hole thermodynamics\cite{eqg} --- see refs.
\cite{hawkpage} and \cite{mads} for discussions relevant for AdS.
There is a problem that must be faced with this approach in that
typically the gravity action diverges. For $d=n{+}1$ spacetime
dimensions, the familiar (Euclidean) action has two contributions
\begin{equation}
I_{\rm bulk}+I_{\rm surf}=-{1\over 16\pi G}\int_{\cal M}
d^{n+1}x\sqrt{g}
\left(R+{n(n-1)\over l^2}\right)
-{1\over 8\pi G}\int_{\partial\cal M} d^{n}x\sqrt{h} K \ .
\labell{classact}
\end{equation}
The first term is just the Einstein--Hilbert--anti--de--Sitter action
with cosmological constant $\Lambda{=}-{n(n-1)\over 2l^2}$.  The
second integral is the Gibbons--Hawking boundary term which is
required so that upon variation with metric fixed at the boundary, the
action yields the Einstein equations \cite{eqg}. Here, $K$ is the
trace of the extrinsic curvature of the boundary $\partial\cal M$ as
embedded in $\cal M$. In the AdS context, both of these expressions
are divergent because the volumes of both $\cal M$ and $\partial\cal
M$ are infinite (and the integrands are nonzero). The traditional
approach to circumventing this problem is to perform a ``background
subtraction''.  That is, one produces a finite result by subtracting
from eqn.~\reef{classact} the contribution of a background reference
spacetime, so that one can compare the properties of the solution of
interest relative to those of the reference state. Note, however, that
this subtraction requires that the asymptotic boundary geometries of
the two solutions can be matched in order to render the surface
contribution finite.\footnote{Again, there is the implicit need for a
regularization procedure with regards to the asymptotic boundary.}
Aside from being a technical nuisance, there are certain cases where
an appropriate reference solution is ambiguous or unknown, {\it e.g.,}
topological black holes~\cite{mann,brillet,vanzo,roberto,birm}, and
Taub--NUT--AdS and Taub--Bolt--AdS \cite{cejm,hawkhunt} --- see
discussion
below.

In the context of the AdS/CFT correspondence, there does not seem to
be room for a background subtraction in, for example,
eqn.\reef{genie}.  Remarkably AdS spacetime offers an alternative
approach! The divergences that arise in eqn.~\reef{classact} are all
proportional to local integrals of the background CFT metric
$\gamma_{ij}$ \cite{edads,math}. Thus these divergences can be
eliminated by extending the regularization procedure for the action
with a ``counterterm subtraction.''  That is eqn.~\reef{classact} is
modified to include the subtraction of a finite set of boundary
integrals (with divergent coefficients) involving curvature scalars
constructed from the background metric $\gamma_{ij}$
\cite{sken}. Recently a remarkable insight was  provided by
ref.~\cite{pervijay} (see also ref.~\cite{coral}): If the counterterms
are expressed in terms of the induced metric $h_{ij}$, rather than
$\gamma_{ij}$, then they naturally appear with the appropriate
divergences, as the volume of the regulator surface grows as it
approaches the boundary of AdS. Thus in the counterterm subtraction
approach, one may produce a finite gravitational action by
supplementing the contributions in eqn.~\reef{classact} with an extra
surface integral
\be
I_{\rm ct}={1\over 8\pi G}\int_{\partial\cal M} d^{n}x
\sqrt{h}\,F(l,\cR,\nabla\cR)\ ,
\labell{ctact}
\end{equation}
where the counterterms depend only on the curvature $\cR$ (and its
derivatives) of the induced boundary metric $h_{ij}$ --- see section
III for explicit expressions. That this construction is unique to
asymptotically AdS spaces is apparent because the AdS curvature scale
$l$ is essential in defining the counterterms\footnote{We are excluding
non--polynomial terms, which could be introduced in the absence of 
a cosmological constant \cite{lau}.} Note that these
expressions are universal depending only on $l$ and the spacetime
dimension. Once these are fixed, one may use the same counterterms to
regulate the action for any choice of coordinates on any
asymptotically AdS solution.\footnote{Actually this is not quite the
complete story --- see below, and section VI.}

Even outside of the AdS/CFT correspondence, counterterm subtraction
provides a remarkable new theoretical tool with which to investigate
gravitational physics. Together, eqns.~\reef{classact} and \reef{ctact}
provide a finite covariant definition of the gravitational action for
asymptotically AdS spaces. As a simple example, one might consider the
energy of a gravitating system in AdS space. Traditionally the
definition of energy in gravity required a background reference
solution in asymptotically AdS spaces\cite{mads,adhh}, just as in
asymptotically flat spaces\cite{adm}. Combined with the quasilocal
formulation of Brown and York \cite{bryk}, the AdS action with
counterterms provides a definition of energy that is independent of
any reference solution
\cite{pervijay}. Using this technique, one discovers a finite energy
for AdS$_5$ with an $\IR{\times}S^3$ boundary. In the context of the
AdS/CFT correspondence, one can interpret this energy as the Casimir
energy of the dual field theory in the latter background geometry
\cite{pervijay}. A similar Casimir energy arises in AdS$_3$
\cite{pervijay}, where
there is a well known difference between the energy $M{=}-1/(8G)$ of
global AdS$_3$ and that of the $M{=}0$ state (which is only locally
AdS$_3$).

Thus one might revisit Euclidean quantum gravity with this new
theoretical tool in hand. In particular, one can address the cases
where the background subtraction technique was not possible or (due to
ambiguiuties) the results were disputed. This is one of the primary
objectives of the current investigation.

The issue of the correct reference state has been disputed for
``topological black holes'' \cite{mann,brillet,vanzo,roberto,birm}, in
particular for the ``hyperbolic AdS black holes''.  The latter are
black hole solutions where the horizon is a hyperbolic space $H^n$
instead of a sphere.  As it happens, there is among these solutions
one which is locally (though not globally) equal to AdS. However, in
order for it to be regular, the Euclidean time has to take a fixed
finite value --- in other words, it is a finite temperature solution.
As such, it is not an adequate reference state for matching
calculations, which would require a solution that admits arbitrary
Euclidean period. In section IV, we apply the counterterm subtraction
prescription to compute the action, and discover some intriguing
results.  We are led to speculate on a connection to the ``precusor''
states recently discussed in ref.~\cite{joelenny}.

The Taub--NUT solution is known to admit an extension to include a
cosmological constant\cite{kramer}, and as such, the
Taub--NUT--anti--de--Sitter (TN--AdS) solution has been studied
recently in refs.\cite{cejm,hawkhunt}. The boundary geometry can not
be matched to that of AdS space, and so there is no known reference
solution with which to make a background subtraction. Instead, in
refs.\cite{cejm,hawkhunt}, the analog of the self--dual TN solution
({\it i.e.,} the one with a ``nut'', a zero--dimensional fixed point
set, at its origin) was used as the reference state in a background
subtraction calculation of the action of the
Taub--Bolt--anti--de--Sitter (TB--AdS) solution.  In section V, we use
the counterterm subtraction for a backgroundless calculation of the
action of TN--AdS. This allows us to study the thermodynamics of this
solution in and of itself.  In particular, we can study its local
intrinsic stability, and find its entropy, as a function of the nut
charge.  This leads to some surprises.

As mentioned above, the counterterm subtraction approach can not be
extended in a straightforward way to asymptotically flat (AF)
spacetimes (and for that matter, to spacetimes which do not asymptote
to AdS) because the AdS scale is an essential ingredient in the
definition of the counterterms \reef{ctact}. However, one can apply
this technique in a case--by--case manner to the computation of the
action of those asymptotically flat solutions which can obtained as
limits of AdS solutions.  A simple example is the computation of the
action of the Schwarzschild solution by first embedding it in AdS.
There exists a Schwarzschild--AdS solution\cite{hawkpage} ---
discussed extensively in the context of the AdS/CFT correspondence
recently \cite{edads,edadsii} --- which for black holes that are much
smaller than the cosmological length scale $l{\sim}|\Lambda|^{-1/2}$,
approximates the asymptotically flat Schwarzschild solution.  We can
compute the action of this Schwarzschild--AdS black hole by using the
counterterm prescription, and then take the limit
$l{\rightarrow}\infty$.  In this way we almost recover the standard
result
that is obtained by matching the AF solution to Minkowski
spacetime.

The preceding is a satisfying, but somewhat trivial example.  However,
there are other cases of AF spaces where the computation of the
action, using the more traditional background subtraction technique,
has been the subject of some controversy. One such case is that of the
Taub--NUT solutions, which are only asymptotically locally flat (ALF).
In ref.\cite{malcolm} the action of generic Euclidean Taub--NUT
solutions (of which only the self--dual Taub--NUT and Taub--bolt
instantons are regular) was computed by trying to match the solutions
to Minkowski space, in order to perform a regularizing subtraction (a
similar matching was also attempted in ref.\cite{gauntetal}).
However, since the large radius slices of Euclidean Taub--NUT space
are squashed three--spheres, in contrast to the Minkowskian slices
$S^1{\times}S^2$, the matching is not really well defined.  Therefore,
it was proposed in ref.\cite{hunter} that the proper background to be
subtracted is instead the self--dual Taub--NUT instanton, which has
the lowest possible energy among the regular Euclidean Taub--NUT
solutions---the only other regular solution is the Taub--bolt
instanton\footnote{Some care should be exercised, since often in the
literature the name ``Taub--NUT solution'' is used to refer
specifically to the self--dual Taub--NUT instanton, instead of the
full, two--parameter Taub--NUT solution.}.  It was noticed in
ref.~\cite{cejm} that there existed a branch of solutions which tends
to the ALF Taub--NUT solution as $l{\rightarrow}\infty$. (These are
the analogue of the ``small'' Schwarzschild black hole branch of
solutions on ref.\cite{hawkpage}.)  Therefore, after applying the
counterterm subtraction procedure to compute the action of the
asymptotically TN--AdS solution, we take the limit
$l{\rightarrow}\infty$.  This limit provides then a ``background
independent'' result for the action of the ALF Taub--NUT solutions.
Remarkably, we find that the result agrees precisely with the
``imperfect matching'' one given in ref.\cite{malcolm}.  Furthermore,
we show that the counterterm prescription results are reproduced by
performing an ``imperfect matching'' to AdS similar to the one in the
ALF case.

A simple application of the counterterm subtraction is to calculate
the action of (Euclidean) AdS$_{n+1}$ for different choices of
coordinates, {\it i.e.}, with different boundary geometries. In
section VI, we present an analysis of the multi--slicing phenomenon
for (Euclidean) AdS$_{n+1}$ with $n{\leq}4$, showing the results for
the action in several different cases.  It is interesting to note the
appearance of different Casimir energies in the various cases.  A more
dramatic result is that for certain boundary geometries, such as $S^n$
and $H^n$, one finds that the counterterm subtraction is
insufficient. That is, a divergence that is logarithmic in the
asymptotic radius appears, and can not be eliminated by the addition
of a local counterterm as in eqn.~\reef{ctact}. These divergences
which can arise for even $n$ have been noted previously in the context
of the AdS/CFT correspondence \cite{edads,sken}. There they may be
related to a conformal anomaly for the dual CFT in certain background
geometries which is well known to be connected to the appearance of
logarithmic divergences in the effective field theory action
\cite{birdav}. Of course, this presents a limitation on counterterm
subtraction as a general tool to investigate asymptotically AdS spaces
in odd spacetime dimensions.

Certainly our results have many interesting implications for the
dual field theory via the AdS/CFT correspondence. We will
only make limited comments on this aspect of the work here, leaving
a more general study of the field theoretic interpretation
for a future paper.

While this work was being completed, we were informed that
R. Mann\cite{robnut} had also considered the application of the AdS
action with counterterms to the solutions considered in sections IV
and V.

\section{The many faces of AdS}\label{sec:faces}

As described in the previous section, counterterm subtraction works by
subtracting the integral of various boundary curvature invariants
\reef{ctact} from the standard action \reef{classact}.  This leaves
unspecified the way in which the boundary of AdS is approached, {\it
i.e.}, the choice of ``radial" coordinate defining the family of
surfaces which approach the boundary as a limit. Depending on this
choice, the slices at constant radius can have different geometry or
even different topology. Even if the spaces are locally equivalent to
one another, the computation of the action will in general lead to
different results, since the boundary terms in the action will take
different values. In the end, since all different forms of a
spacetime will be related by diffeomorphisms, with possible addition
or subtraction of points, and possibly as well, identifications under
discrete subgroups, the different results for the action will bear a
relation to one another, too. Here we will describe some of the many
possible ``faces" of the boundary of AdS. In subsequent sections, we
will consider these metrics as examples for the application of the
counterterm subtraction technique, and compare the results. Clearly,
such a comparison would have been impossible had we required a
background for the calculation.

Let us first present Euclidean AdS$_{n+1}$ in the following three
familiar
metrics,
\be
ds^2=\left( k+{r^2\over l^2}\right)d\tau^2 +{dr^2\over \left(
k+{r^2\over
l^2}\right)} +{r^2\over l^2} d\Sigma^2_{k,n-1}\ ,
\labell{metric1}
\ee
where the $(n{-}1)$--dimensional metric $d\Sigma^2_{k,n-1}$ is
\begin{equation}
d\Sigma^2_{k,n-1} =\left\{ \begin{array}{ll}
\vphantom{\sum_{i=1}^{n-1}}
l^2d\Omega^2_{n-1}& {\rm for}\; k = +1\\
\sum_{i=1}^{n-1} dx_i^2&{\rm for}\; k = 0 \\
\vphantom{\sum_{i=1}^{n-1}}
l^2d\Xi^2_{n-1} &{\rm for}\; k = -1\ ,
\end{array} \right.
\labell{little}
\end{equation}
where $d\Omega^2_{n-1}$ is the unit metric on $S^{n-1}$. By $H^{n-1}$
we mean the $(n{-}1)$--dimensional hyperbolic space, whose ``unit
metric" $d\Xi^2_{n-1}$ can be obtained by analytic continuation of
that on $S^{n-1}$. It is straightforward to see that all of the above
solutions are locally equivalent to each other. In the above we are
assuming that $n{>}2$ for $k{=}{-}1$, since for $n{=}2$ one does not
have a
hyperbolic metric $H^1$.

For later use in the paper, we will write the volume of the space
$d\Sigma^2_{k,n}$ as $l^n\sigma_{k,n}$. In this way, $\sigma_{k=+1,n}$
will
be equal to the volume $\omega_n$ of the unit $n-$sphere.

Next we consider Euclidean AdS$_{n+1}$ with metric
\be
ds^2={dr^2\over \left( k+{r^2\over
l^2}\right)} +{r^2\over l^2} d\Sigma^2_{k,n}\ ,
\labell{metric2}
\ee
where the $n$--dimensional metric $d\Sigma^2_{k,n}$ is defined
precisely in the same way as above in eqn.~\reef{little}.  For $k{=}0$,
this simply reproduces the $k{=}0$ metric in eqn.~\reef{metric1}.  One
might note that a transformation of the radial coordinate brings these
metrics into the form
\be
ds^2=l^2d\rho^2+f_k^2(\rho) d\Sigma^2_{k,n}\ ,
\labell{metric2b}
\ee
where
\begin{equation}
f_{k}(\rho) =\left\{ \begin{array}{lll}
\sinh\rho& {\rm for}\; k = +1\\
e^\rho& {\rm for}\; k = 0 \\
\cosh\rho&{\rm for}\; k = -1\ .
\end{array} \right.
\labell{radd}
\end{equation}

\noindent
One final AdS metric which we will consider is \be ds^2={dr^2\over
\left( k+{r^2\over l^2}\right)} +\left( k+{r^2\over
l^2}\right)d\widehat\Sigma_{-k,\hat m}^2 +{r^2\over l^2}
d\widetilde\Sigma^2_{k,\tilde m}\ , \labell{metric3} \ee where again
the metrics $d\widehat\Sigma_{-k,\hat m}^2$ and
$d\widetilde\Sigma^2_{k,\tilde m}$ are defined in eqn.~\reef{little}.
For $k{=}0$ we once again reproduce the $k{=}0$ metric in
eqn.~\reef{metric1}.  For $k{=}{\pm}1$, we assume that both $\hat
m,\tilde m\ge2$.  For $k{=}{+}1$, the boundary geometry is $H^{\hat
m}{\times}S^{\tilde m}$, while for $k{=}{-}1$, we simply interchange the
hyperbolic space and the sphere. However, in the latter case, the
coordinate transformation ${\tilde r}^2{=}r^2{-}l^2$ puts the metric
back in the $k{=}{+}1$ form with $\hat m{\leftrightarrow}\tilde m$.

Thus with the metrics in eqns.~\reef{metric1},~\reef{metric2} and
\reef{metric3}, we have displayed AdS$_{n+1}$ with a
wide variety of boundary geometries:
\be
\IR^n,\ S^1\times S^{n-1},\ S^1\times \IR^{n-1},\ S^1\times H^{n-1},\
S^{n},\ H^n,\ S^m\times H^{n-m}\ .
\labell{display}
\ee
All of these AdS metrics are maximally symmetric, {\it i.e.,}
\be
R_{ijkl}=-{1\over l^2}(g_{ik}\,g_{jl}-g_{il}\,g_{jk})\ ,
\labell{maxi}
\ee
which ensures that the geometry is conformally flat.
This condition also ensures the geometries are all locally AdS.

It is interesting to notice the form of some of the boundary geometries
we
get here upon analytic continuation to
Minkowski signature, since they are rather common :
\be
\begin{array}{lll}
S^{n}&:& {\rm Euclidean\; de\; Sitter\; space}\\
H^{n}&:& {\rm (global)\; anti-de\; Sitter\; space}\\
\IR^{n}&:& {\rm Minkowski\; space\;}.\\
\end{array}
\ee
Furthermore, if we assume a specific analytic continuation to Lorentzian
spacetime,
{\it e.g.}, $S^1{\times}S^n \rightarrow  \IR({\rm time}){\times}S^n$,
then
\be
\begin{array}{lll}
\IR\times S^{n-1}&:& {\rm the\; Einstein\; static\; universe\;} \\
\IR\times H^{n-1}&:& {\rm the\; static\; open\; universe\;}.\\
\end{array}
\ee
The AdS/CFT correspondence implies then an equivalence between, on the
one hand, quantum gravity in AdS and, on the other hand, a CFT on any 
of the above backgrounds. We find it particularly amusing that, when 
the boundary
is taken to be $H^n$, quantum gravity in AdS$_{n+1}$ can be dual to a
CFT on an AdS$_n$ background! It should be kept in mind that the 
geometry on the boundary is not dynamical, since there are no 
gravitational degrees of freedom in the dual CFT.

There is an important feature that distinguishes the solutions with
$k{=}{-}1$ from those with $k{=}0,{+}1$: there is a finite minimum
radius $r{=}l$ at which $g_{rr}$ diverges. In eqn. \reef{metric1}, the
Killing vector $\partial_\tau$ also has fixed point set (a ``bolt") at
this radius. In this case, the Euclidean solution will be regular only
if the coordinate $\tau$ is identified with period $\beta{=}2\pi l$.
In the metric \reef{metric2} with $k{=}{-}1$, the minimum radius
$r{=}r_+$ simply denotes the boundary of a coordinate patch as is
evident from the form of the metric in eqn.~\reef{metric2b} with the
new radial coordinate $\rho$. In the case of eqn.~\reef{metric3} with
$k{=}{-}1$, $r{=}r_+$ is the location of a ``conical'' singularity.
For $k{=}{+}1$, the minimum radius is $r{=}r_+{=}0$ and the geometry
is smooth at this point in metrics
\reef{metric1} and \reef{metric2}, but it corresponds to a ``conical
singularity'' in eqn.~\reef{metric3}.  For $k{=}0$, the minimum radius
is again $r{=}r_+{=}0$ which in this case is an infinite proper
distance away and so there is no problem with the curvature here.
Note that the geometries with conical singularities or a bolt are only
locally AdS, that is they describe AdS spacetime with additional
discrete identifications of points.

Eqn.~\reef{maxi} is an extremely restrictive condition.
If one is simply interested in solving Einstein's equations with
a negative cosmological constant
\be
R_{ij}=-{n-2\over l^2}g_{ij}\ ,
\ee
then the above metrics remain solutions when the boundary geometries
are replaced by arbitrary Einstein spaces. In all of the metrics
(\ref{metric1},\ref{metric2},\ref{metric3}), one may replace any of
the $S^p$ factors (with $p{>}1$) by a space satisfying $\widetilde
R_{ab}{=}(p-1)/l^2\,\tilde g_{ab}$. Similarly any $H^p$ factors can be
replaced by a space satisfying $\widehat R_{ab}{=}{-}(p-1)/l^2\,\hat
g_{ab}$, and $\IR^p$ factors can be replaced by any Ricci flat
solution, {\it i.e.,} $R_{ab}{=}0$. For example, then $S^p$ can be
replaced by
a product of spheres $S^{p_1}{\times}\cdots{\times}S^{p_q}$ where
$\sum_{i=1}^qp_i{=}p$ with $p_i{>}2$ and the radii of the individual
spheres is scaled so $r_i^2{=}(p-1)/(p_i-1)\,l^2$. These generalized
solutions will no longer be conformally flat or locally
AdS. Furthermore generically a true curvature singularity is
introduced at the minimum radius, {\it e.g.}, $R_{ijkl}R^{ijkl}$ grows
without bound as $r$ approaches $r_+$.

\section{Counterterm action}

The detailed form of the boundary counterterms was originally explored
in ref.~\cite{sken}, where they were derived in terms of the
background (field theory) metric $\gamma_{ij}$. The insight provided
by ref.~\cite{pervijay} was that the counterterms should be written in
terms of the induced metric on the boundary $h_{ij}$. In this way,
they naturally appear with the appropriate (infinite volume)
divergences to cancel those arising from the classical gravitational
action.  The focus of ref.~\cite{pervijay} was to construct a
finite boundary stress--tensor without using a reference background.
However, the proposed prescription naturally provides the construction
of a finite action which can then be employed, for example, to
calculate the action of Euclidean gravitational instantons. This will
be the primary application which we consider in the following.

Hence the full (Euclidean) gravitational action in $d{=}n{+}1$ spacetime
dimensions has three contributions
\begin{equation}
I_{\rm AdS}=I_{\rm bulk}(g_{ij})+I_{\rm surf}(g_{ij})+I_{\rm
ct}(h_{ij})\ .
\labell{total}
\end{equation}
The first two terms, comprising the familiar classical action, were
given in eqn.~\reef{classact}. Here, $h_{ij}$ is the induced metric on
the boundary $\partial\cal M$ which may be defined as
$h_{ij}{=}g_{ij}{-}n_in_j$ where $n$ is an outward pointing unit normal
vector to $\partial\cal M$. In the Gibbons-Hawking boundary term
$I_{\rm surf}$, the trace of the extrinsic curvature is defined by
$K{=}h^{ij}\nabla_in_j$.\footnote{Our conventions differ by signs from
refs.~\cite{pervijay,bryk}, but are chosen to conform with standard
practice in General Relativity, as in, {\it e.g.},
ref.\cite{waldbook}.}

The counterterm action $I_{\rm ct}(h_{ij})$ may be arranged as an expansion
in powers of the boundary curvature (and its derivatives). The number
of terms that appears grows with the dimension of the spacetime.
The first few terms are explicitly
\begin{equation}
I_{\rm ct}={1\over 8\pi G }\int_{\partial\cal M} d^{n}x\sqrt{h}
\left[{n-1\over l}+{l\over 2(n-2)}{\cal R}+
{l^3\over2(n-4)(n-2)^2}\left({\cal R}_{ab}{\cal R}^{ab}-{n\over4(n-1)}
{\cal R}^2\right)+
\dots\right]\ ,
\labell{explete}
\end{equation}
where $\cal R$ and ${\cal R}_{ab}$ are the Ricci scalar and Ricci
tensor for the boundary metric, respectively. Combined these three
counterterms are sufficient\footnote{Or almost, see Section~IV.} to
cancel divergences for $n\leq 6$.  In this covariant form, the first
term originally appeared in ref.~\cite{coral}\footnote{This term had
also been considered to provide a (partial) regularization of the
action of AdS$_5$ in ref.\cite{liutse}.}, while the second term first
appeared in ref.~\cite{pervijay}. We derived the third term by
demanding that the infinite volume divergences were cancelled when
using the metric \reef{metric3}. Any of these terms may be derived
with the construction provided by ref.~\cite{sken} for the appropriate
curvature integral in terms of the CFT metric $\gamma_{ij}$.  One then
simply substitutes the induced boundary metric $h_{ij}$ to produce the
covariant counterterms appearing in $I_{\rm ct}$.  To go to higher
dimensions, resorting to this construction seems inescapable as the
``simple'' asymptotically AdS metrics presented in section II can not
be used to distinguish all of the curvature invariants that can appear
in the higher order counterterms.  It is important to note that the
fact that we have counterterms for dimensions up to $d{=}7$ means that
we can now study all (known) AdS applications which arise in string
theory and M--theory.

Other matter field actions, for example an action for Maxwell fields,
can be added to eqn.~\reef{total}.  Although, at least for black hole
solutions, the addition of gauge fields does not seem to require new
counterterms\cite{cejm2}, we must remain alert to the possibility that
extra matter fields may require the addition of new, non--geometric
surface counterterms to the action. This issue will not be considered
further here.

 As a simple example, we will consider calculating the action
\reef{total}
with the metric \reef{metric1} for AdS with boundary
$S^1{\times}M_k$. Let us present the contributions of the individual
terms in the action:
\bea
I_{\rm bulk}&=&{\beta\sigma_{k,n-1}\over 8\pi Gl^{2}}
\left[-r_+^n +r^n\right]\ ,
\labell{contire}\\
I_{\rm surf}&=&{\beta\sigma_{k,n-1}\over 8\pi Gl^{2}}
\left[r^n\left(-n-k(n-1){l^2\over r^2}\right)\right]\ ,
\nonumber\\
I^1_{\rm ct}&=&{\beta\sigma_{k,n-1}\over 8\pi Gl^{2}}
\left[r^n(n-1)\left(1+k{l^2\over r^2}\right)^{1/2}\right]
\nonumber\\
&=&{\beta\sigma_{k,n-1}\over 8\pi Gl^{2}}
\left[r^n(n-1)\left(1+{k\over2}{l^2\over r^2}-{k^2\over8}{l^4\over r^4}
+{k^3\over16}{l^6\over r^6}+\cdots\right)\right]\ ,
\nonumber\\
I^2_{\rm ct}&=&{\beta\sigma_{k,n-1}\over 8\pi Gl^{2}}
\left[r^n(n-1){k\over2}{l^2\over r^2}\left(1+k{l^2\over
r^2}\right)^{1/2}
\right]
\nonumber\\
&=&{\beta\sigma_{k,n-1}\over 8\pi Gl^{2}}
\left[r^n(n-1)\left(\ \ {k\over2}{l^2\over r^2}+{k^2\over4}{l^4\over
r^4}
-{k^3\over16}{l^6\over r^6}+\cdots\right)\right]\ ,
\nonumber\\
I^3_{\rm ct}&=&{\beta\sigma_{k,n-1}\over 8\pi Gl^{2}}
\left[r^n(n-1)\left(-{k^2\over8}{l^4\over r^4}\right)
\left(1+k{l^2\over r^2}\right)^{1/2}
\right]
\nonumber\\
&=&{\beta\sigma_{k,n-1}\over 8\pi Gl^{2}}
\left[r^n(n-1)\left(\ \ \ \ -{k^2\over8}{l^4\over r^4}
-{k^3\over16}{l^6\over r^6}+\cdots\right)\right]\ ,
\nonumber
\eea
where $\sigma_{k,n-1}$ is the (dimensionless) volume of the space
with metric $d\Sigma^2_{k,n-1}/l^2$, and
$\beta$ is the period of $\tau$. We have also separated the
contributions of the individual counterterms in \reef{explete}, so
$I^i_{\rm ct}$ is the integral of the $i$'th term in the action.
Now, for a particular boundary dimension only some of the counterterms
are included to cancel the divergences. So for $n{=}2i{-}1,2i$, one
keeps
only up to $I^i_{\rm ct}$.  For any odd value of $n$, one has then
\begin{equation}
I_{k,n+1}=-{\beta\sigma_{k,n-1}\over8\pi Gl^{2}}
\left(r_+^n+O(l^{n+1}/r)\right)\ .
\labell{odd}
\end{equation}
For the even values of $n$, an extra constant term makes an
appearance so that
\be
I_{k,n+1}={\beta\sigma_{k,n-1}\over8\pi Gl^{2}}\left(-r_+^n
-{k\over 2}l^2\delta_{n,2}+{3k^2\over
8}l^4\delta_{n,4}-{5k\over 16}l^6\delta_{n,6}+\dots+O(l^{n+1}/r)\right)\
.
\labell{even}
\ee
As we have explained above, for $k{=}{+}1,0$, we have $\beta$
arbitrary and $r_+{=}0$, whereas for $k{=}{-}1$, $r_+{=}l$ and
$\beta{=}2\pi l$.

Note that for even $n$, the coefficients of the higher counter terms
are actually divergent, even though they formally evaluate to a finite
result.  Further in either of these results, eqns.~\reef{odd} and
\reef{even}, there are extra terms of order $1/r$, which vanish when
the limit $r{\rightarrow}\infty$ is taken in order to approach the AdS
boundary. However consider the case of $n$ odd, where we have in fact
the option of keeping all of the higher order counter terms in
eqn.~\reef{explete}, {\it i.e.}, including the terms which actually
vanish in the boundary limit. This would give a result where in fact
all of the inverse powers $r^{-p}$ would be cancelled so that not only
would the action be finite, but it would be independent of the
regulator radius!

Given the explicit counterterms in eqn.~\reef{explete},
we can only really evaluate the action for $n{\le}6$. However,
keeping in mind that the higher order counterterms ensure the
cancellation of divergences order by order, it is clear that the
formulae \reef{odd} and \reef{even} will be unchanged for $n>6$.
Further we can show that the coefficient of the extra contributions
for $n$ even will be
\be
(-k)^{n/2} {(n-1)!!^2\over n!} l^n\ .
\labell{extra}
\ee
To derive this result, note that the bulk and surface contributions
can be written as
\be
I_{\rm bulk}+I_{\rm surf}= {\beta\sigma_{k,n-1}\over 8\pi Gl^2}\left(
-r^n_+-(n-1)r^n(1+x)\right)\ ,
\label{lazy}
\ee
where $x{=}kl^2/r^2$, while the counterterms yield
\be
I^p_{\rm ct}={\beta\sigma_{k,n-1}\over 8\pi Gl^2}(n-1)r^n\, c_px^{p-1}\,
(1+x)^{1/2}
\label{lazya}
\ee
where $c_p$ are constants independent of $n$. The key point is to
realize that the counterterm contributions will cancel the $x$--dependence
in eqn.~(\ref{lazy}) to an arbitrarily large order, and hence these
coefficients are just the coefficients in the Taylor series
\be
(1+x)^{1/2}=\sum_{p=1}^\infty c_p x^{p-1}\ .
\label{lazyb}
\ee
Now as stated above for a given $n{=}2i$, the action  only includes
the finite sum: $I_{\rm bulk}+I_{\rm surf}+\sum_{p=1}^iI^p_{\rm ct}$.
Thus with some elementary manipulations, one finds the residual finite
term in eqn.~(\ref{even}) appears with the coefficient (\ref{extra}) above.

\section{AdS Black Holes}\label{sec:hyper}

In this section we turn to the study of black hole solutions, using
the counterterm subtraction scheme.  In the presence of a negative
cosmological constant, the horizon of a black hole admits a much
larger variety of geometries and topologies than in asymptotically
flat situations. This is consistent with the variety of boundary
topologies that we can obtain for AdS itself, depending upon how we
choose to radially foliate it, as discussed in
section~\ref{sec:faces}. The case ($k{=}1$, below) of spherical black
holes has already been studied using this counterterm subtraction
scheme in ref.~\cite{pervijay}, but we compute and list those results
in what follows for completeness and for comparison with the flat and
hyperbolic cases.

In ref.\cite{birm}, it was shown that the Einstein--anti---de--Sitter
system in $n{+}1$ dimensions admits the following solutions: \be
ds^2=-V_k(r)dt^2 +{dr^2\over V_k(r)}+{r^2\over l^2} d\Sigma^2_{k,n-1}\
, \ee with \be V_k(r) =k-{\mu\over r^{n-2}}+{r^2\over l^2}\ , \ee
where the $(n{-}1)$ dimensional metric $d\Sigma^2_{k,n-1}$ is defined
as in eqn.~\reef{little}. Thus it represents $S^{n-1}$, $\IR^{n-1}$
and $H^{n-1}$ for $k=+1,0$ and $-1$, respectively.  A spacetime that
is locally the same as anti--de Sitter is recovered when $\mu{=}0$ for
which the metric reduces to that in eqn.~\reef{metric1}.

By going to the Euclidean section one finds that the Euclidean time
period
(the inverse temperature) has to be
\be
\beta={4\pi l^2 r_+\over nr_+^2+ k(n-2)l^2}\ .
\ee
Here, $r_+$ is the largest positive root of $V_k(r)$, typically
associated to the outer horizon of a black hole. For $k{=}1$ and
$\mu{=}0$ (global AdS), there is no such root, but the correct results
are obtained by setting $r_+{=}0$.  Now, it is important to notice
that, whereas for $k{=}\{1,0\}$, the locally AdS solution corresponds
to $r_+{=}0$, this is not true for $k{=}-1$. AdS spacetime with hyperbolic
slicing has a bifurcate Killing horizon at $r{=}l$, and a fixed
temperature $\beta{=}2\pi l$. By contrast, there exists an extremal
$k{=}{-}1$ solution, with a degenerate horizon at $r{=}r_e$ and
parameter
$\mu{=}\mu_e$, satisfying
\be
r_e=\sqrt{n-2\over n} l\ ,\qquad \mu_e = -{2\over n-2}\left({n-2\over
n}\right)^{n/2} l^{n-2}.
\ee
In particular,
\be\label{ext3}
\mu_e=-{2l\over 3\sqrt{3}}\ ,\qquad r_e={l\over\sqrt{3}},\qquad {\rm
for\;}
n=3\ ,
\ee
\be\label{ext4}
\mu_e=-{l^2\over 4}\ ,\qquad r_e={l\over\sqrt{2}},\qquad {\rm for\;}
n=4\ ,
\ee
\be\label{ext6}
\mu_e=-{4l^4\over 27}\ ,\qquad r_e=\sqrt{2\over 3}l,\qquad {\rm for\;}
n=6\ .
\ee
Therefore, in a calculation for $k{=}{-}1$ of the action, with
background matching, the question arises concerning which is the
correct background to subtract: On the one hand, the locally AdS
solution ---which has the higher symmetry--- might be physically
appealing. However, since its period $\beta$ is fixed, matching it to
a solution with a different value of~$\beta$ would introduce a conical
singularity at the horizon\cite{vanzo}. On the other hand, the
extremal solution, with a lower value of~$\mu$ (and as we will see, of
the energy), has arbitrary $\beta$ and therefore can be matched to any
other solution. Hence, the extremal solution was the preferred
background for the matching calculations in
refs.\cite{vanzo,roberto,birm}.

It is clear from this discussion that the method of counterterm
subtraction can be of help here. For the solutions described above we
obtain:
\bea\label{hypact}
I_{k,n+1}&=&{\beta\sigma_{k,n-1}\over 8\pi G l^{2}}\left( -r_+^n +{\mu
l^2\over 2} -{k\over 2}l^2\delta_{n,2}+{3k^2\over
8}l^4\delta_{n,4}-{5k\over 16}l^6\delta_{n,6}+\dots \right)\nonumber\\
&=&{\beta\sigma_{k,n-1}\over 16\pi G l^{2}}\left(kr_+^{n-2}l^2 - r_+^n
-{k}l^2\delta_{n,2}+{3k^2\over 4}l^4\delta_{n,4}-{5k\over
8}l^6\delta_{n,6}+\dots\right)\ ,
\eea
where again $\sigma_{k,n-1}$ is the (dimensionless) volume associated
with the unit metric $d\Sigma^2_{k,n-1}/l^2$. Using
eqn.(\ref{hypact}) we can compute the energy and entropy of the
solutions by application of standard thermodynamical formulas.  One
finds
\be\label{hypen}
E={(n-1)\sigma_{k,n-1}\over 16\pi G}\mu+E_k^0\ ,
\ee
where we denote by
\be
E_k^0={\sigma_{k,n-1}\over 16\pi G}\left(-{k}\delta_{n,2}+{3k^2\over
4}l^2\delta_{n,4}-{5k\over 8}l^4\delta_{n,6}+\dots \right)\ ,
\ee
the terms that are independent of the black hole parameters ({\it
e.g.}, of the temperature). Their contribution to the action is
therefore of the form~$\beta E_k^0$. Note that one can
extrapolate this Casimir energy to
\be
E_k^0={\sigma_{k,n-1}\over 8\pi G}(-k)^{n/2}{(n-1)!!^2\over n!} l^{n-2}\
,
\labell{zero}
\ee
for arbitrary even $n$ using eqn.~\reef{extra}.

The entropy,
\be
S= {\sigma_{k,n-1}r_+^{n-1}\over 4G}\ ,
\ee
satisfies the area law, and is independent of the extra terms $\beta
E_k^0$.  Not surprisingly, the result is therefore the same as in a
background calculation.

Curiously, the results for $n{=}3$ and $n{=}4$ show different
qualitative features. For $n{=}3$ the result that we obtain is the
same as one would obtain by performing a background subtraction from
the locally AdS$_4$ solution {\it neglecting the conical singularity
that would appear for $k{=}{-}1$}. This is rather similar to what we
will find for TN--AdS in the next section:
the method of counterterm subtraction appears to
reproduce the results of an ``imperfect matching'' calculation.  As a
result, the extremal solution (\ref{ext3}) has negative energy,
whereas the locally AdS solution, with $\mu{=}0$, has vanishing action
and energy.

By contrast, the result for the hyperbolic $n{=}4$ black holes
supports the opposite scenario! The action (\ref{hypact}) in this case
reproduces precisely that obtained by taking the extremal state
(\ref{ext4}) as the reference state, and not the locally AdS state
(notice that $I{=}0$ for the values in eqn.~(\ref{ext4})). For $n{=}4$
and
$k{=}{-}1$, the energy (\ref{hypen}) of the extremal state vanishes, a
confirmation that this is to be taken as the ground state of the
theory.  The term $E_k^0$ is independent of the black hole parameters
({\it e.g.} the temperature), and its contribution to the action is
therefore simply of the form~$\beta E_k^0$.

For $k{=}1$ this term has been identified in ref.\cite{pervijay} as
precisely the Casimir energy associated to ${\cal N}{=}4$
super--Yang--Mills theory on the static Einstein spacetime
$\IR{\times}S^3$, which is the spacetime obtained as the boundary of
AdS in this case. This agreement is a striking outcome of the
counterterm subtraction method. Notice that the interpretation as a
Casimir energy is the only possible one, given that the AdS solution
is the one with the lowest action and energy among that family---{\it
i.e.}, it is the ground state.

We would like to see whether a similar correspondence holds for
$k{=}{-}1$. In this case it is crucial to notice that the ground state
is {\it not} the locally AdS solution. The latter should be regarded
as an excited state of the system. The ground state is the extremal
solution, which has zero energy. By translating this into the AdS/CFT
correspondence we would not expect to find a Casimir energy for the
field theory calculations on the open static universe
$\IR{\times}H^3$. Indeed, the effective action and renormalized
stress--energy tensor for conformal fields vanish on that space (see,
{\it e.g.,} ref.\cite{birdav}). This is in perfect agreement with the
zero energy results that we find for the ground state (\ref{ext4}) of
the theory.

There are, however, some aspects that are in need of further
exploration.  In particular, from the entropy formula we see that for
$k{=}{-}1$, not only the does locally AdS solution have non--zero
entropy,
but
so does the extremal ground state. In particular, for $n{=}4$,
\be\label{extent}
S_{\rm ext}={ \sigma_{k=-1,3}l^3\over 2^{7/2} G}\ .
\ee
In this respect, this ground state bears resemblance to the extremal
black hole ground state state discussed in \cite{cejm2}, which had
non--vanishing entropy as well. It is of great interest to understand
this result (\ref{extent}) from a field--theoretical point of view.
The ``precursor'' states of ref.\cite{joelenny} ---constructed in
standard field theory--- might be extremely relevant to such a
discussion. As proposed in ref.\cite{joelenny}, these are degrees of
freedom that do not contribute to the energy density, although they
store information. This looks precisely like what is needed to account
for an entropy like we have found in eqn.~(\ref{extent}). Perhaps the
entropy of this ground state and the one presented in ref.\cite{cejm2}
represents the count of the number of precursor degrees of freedom in
the field theory.

For black holes in AdS$_6$ ({\it i.e.}, $n{=}5$, and in fact, all odd
values of $n$) the conclusions are essentially the same as in AdS$_4$.
However, the situation for AdS$_7$ is somewhat enigmatic.  In this
case, the action does not vanish either for the extremal black hole or
for the locally AdS solution!  Also, the energy is non--zero for both.
Perhaps this is consistent with yet to be understood properties of the
$(2,0)$ superconformal field theory that resides on the world--volume
of the M5--brane\cite{exotic2}.

Finally, it is of interest to note that because the ``small''
Schwarzschild black holes (in the sense of ref.\cite{hawkpage})
survive the $l{\to}\infty$ limit, ({\it i.e.,} the cosmological
constant goes to zero), the surface counterterm subtraction method
supplies results for the action, energy and entropy for ordinary
Schwarzschild black holes. For odd $n$, these results coincide
precisely with those obtained by the background subtraction method,
using Minkowski spacetime as a reference. For even $n$, the results
would again coincide with the standard results in asymptotically flat
space, except for the constant contribution of the Casimir energy
\reef{zero} (and the analogous term in the action).  In this case
because for $n\ge4$ this energy is proportional to $l^{n-2}$, it
becomes an infinite constant in the limit $\l{\rightarrow}\infty$.  We
will see that this ability to take the flat spacetime limit occurs for
other interesting solutions in the next section, and allows us to
address and resolve certain situations which were fraught with
uncertainties and/or ambiguiuties in the literature.

\section{The Anti--de Sitter NUTcracker}

As we mentioned in the introduction, the issue of choosing a correct
reference state for background subtraction has been a matter of some
controversy for Taub--NUT and Taub--bolt solutions, in the
asymptotically locally flat situation \cite{malcolm,hunter} as well as
in the asymptotically locally AdS case \cite{cejm,hawkhunt}.

Note that in this section $n$ will be used to denote the ``nut
charge", not the number of dimensions---we will only deal with
four--dimensional solutions.

\subsection{Spherical nuts and bolts}
The Taub--NUT--Anti--de--Sitter (TN--AdS) solution is
\begin{equation}
ds^2 = V(r) (d\tau + 2n\cos\theta d\varphi)^2
+ {dr^2\over V(r)} + (r^2 -n^2)(d\theta^2+\sin^2\theta d\varphi^2)\ ,
\end{equation}
where
\begin{equation}
V = { (r^2 +n^2)- 2 mr +l^{-2}( r^4 - 6 n^2 r^2-3 n^4)\over r^2 - n^2}\
.
\end{equation}
Here we
will simply sketch some of the features of the solution. For a
detailed analysis we refer the reader to ref.~\cite{cejm}. If $n{=}0$ we
recover the Schwarzschild--AdS solutions with $m$ as a
mass parameter. The analytically continued time, $\tau$, parameterizes
a circle, $S^1$, which is fibred over the two sphere $S^2$, with
coordinates $\theta$ and $\varphi$.  The non--trivial fibration is a
result of a non--vanishing ``nut charge'' $n$. As a result, the
boundary as $r{\rightarrow}\infty$ is described as a ``squashed''
three--sphere, where $4n^2/l^2$ parameterizes the squashing.

Euclidean regularity of the solution restricts the period of $\tau$ to
be
\begin{equation}
\beta{=}8\pi n\ .
\end{equation}
In addition, the mass parameter has to be restricted so that the fixed
point set of the Killing vector $\partial_\tau$ at radial position
$r{=}r_+$ is a regular one.  Hence one finds ``nut'' or ``bolt''
solutions, depending on whether the fixed point set is zero or
two dimensional, respectively. In particular, for ``nut'' solutions
\begin{equation}\label{nutvalues}
r_+=n,\qquad m_{n}=n-{4n^3\over l^2}\ .
\end{equation}
In what follows, by TN-AdS we will mean the Taub--NUT--AdS solutions
with this particular value of $m$. Notice that $m_n$ vanishes for the
value $n{=}l/2$. It was shown in ref.\cite{cejm} that for this
particular value the solution is precisely AdS$_4$, with the slicing
in which the sections at constant $r$ are round three--spheres. In
contrast, the solution with $n{=}m{=}0$ corresponds to AdS$_4$ with
slices of geometry $S^1{\times}S^2$. For Taub--bolt--AdS (TB--AdS) the
expressions are more complicated\cite{cejm}:
\be\label{implies}m_{b}={r_b^2+n^2 \over 2 r_b}
+{1\over 2 l^2} \left(r_b^3 -6n^2 r_b -3{n^4 \over r_b}\right)\ .
\ee
\be\label{boltplace}
r_+=r_{b\pm}=
{l^2 \over 12n} \left(1\pm\sqrt{1-48{n^2\over l^2}+144 {n^4 \over
l^4}}\right)\ .
\ee  For $r_b$ to be real the discriminant must be
non--negative. Furthermore we must take the part of the solution which
corresponds to $r_b{>}n$. This gives:
\be\label{real}
n\leq\left({1\over 6}-{\sqrt{3}\over 12}\right)^{1\over2}l\ .
\ee
It is only for this range of
parameters that one can construct real Euclidean TB--AdS solutions.
Notice, in particular, that the AdS value $l{=}2n$ lies outside this
range.

In refs.\cite{cejm,hawkhunt}, the action of the TB--AdS solutions was
computed by matching the solutions to a TN--AdS solution with the same
value of the nut charge. The thermodynamics of TB--AdS solutions were
then found to be rather similar to that of Schwarzschild--AdS black
holes. However, this method precluded an analysis of the TN--AdS
solutions by themselves, since they acted as reference states.  A
completely rigourous calculation of the action of TN--AdS could not be
performed using the reference background method, simply because it is
not possible to match pure AdS (the intuitively obvious candidate
background) to TN--AdS, as they have incompatible slices for all $n$
except $n{=}l/2$. Equipped with the counterterm subtraction
procedure, we can now compute the action for TN--AdS, without any
reference to a background.

\noindent
With
\begin{equation}
\sqrt{h}= \sqrt{V(r)} (r^2-n^2)\sin\theta\ ,\qquad {\cal R}={2\over
r^2-n^2}-{2n^2\over (r^2-n^2)^2}V(r)\ ,
\end{equation}
we find, for a solution with generic values of $m$ and $n$
\begin{equation}\label{nutact}
I={4\pi n\over Gl^2}\left( l^2 m+3 n^2 r_+-r_+^3\right)\ ,
\end{equation}
where, as we said above, $r_+$ is the minimum possible value of $r$,
where
there is a fixed point of the Killing vector $\partial_\tau$. Of 
course, as explained above, Euclidean regularity demands either $m=m_n$ or 
$m=m_b$.

There are several things to note about this result. The first is a
consistency check: if we subtract the values we obtain for the TB--AdS
and TN--AdS solutions, $I_{\rm bolt}{-}I_{\rm nut}$, we recover (after
some algebra), the result obtained in refs.\cite{cejm,hawkhunt} for
the action of TB--AdS with TN--AdS as a reference. Of course this
consistency is to be expected in general. The standard background
subtraction requires the asymptotic geometry of the solution and its
reference state match. Hence the counterterms which depend only on the
intrinsic boundary geometry must be equal, and will cancel if one
takes the difference of the counterterm subtracted actions.

Next, in the flat space limit $l{\rightarrow}\infty$ we obtain
\begin{equation}\label{garmal}
I\rightarrow {4\pi nm\over G}\ .
\end{equation}
In particular, in this limit we find
\begin{equation}
I_{\rm nut}\rightarrow{4\pi n^2\over G}\ ,\qquad I_{\rm
bolt}\rightarrow{5\pi n^2\over G}\ .
\end{equation}
These are precisely the results that were obtained in
ref.\cite{malcolm} by an ``imperfect match'' of the Taub--NUT solution
to Euclidean Minkowski space.  Indeed, the same ``imperfect match'' to
AdS can be seen to reproduce the result (\ref{nutact}) above. Even if
it is not possible to match the squashed $S^3$ at the boundary to the
boundary of AdS$_4$ with the slicing $S^1{\times}S^2$, a finite result
can nevertheless be obtained by neglecting the non--trivial fibering
and performing a standard background subtraction. Proceeding this way
the bulk (volume) term yields, at large $r$
\begin{equation}\label{nutbulk}
I_{\rm bulk}={4\pi n\over Gl^2}\left( l^2 m+3 n^2 r_+-r_+^3\right) +{\pi
n^3
r\over Gl^2}+ O(1/r)\ .
\end{equation}
In contrast to other action calculations in AdS, the bulk term, even
after
subtraction, is not finite by itself, rather one needs to take into
account the Gibbons--Hawking boundary term:
\begin{equation}\label{nutbdry}
I_{\rm surf}=-{1\over 8\pi G}\int_{\partial\cal M} d^{3}x\sqrt{h}
(K-K_0)=-{\pi n^3 r\over Gl^2}+ O(1/r)\ .
\end{equation}
By adding (\ref{nutbulk}) and (\ref{nutbdry}) and taking
$r\rightarrow\infty$ we therefore recover
eqn.~(\ref{nutact}).

{\it We therefore conclude that the fact that the match to the
background is an imperfect one does not appear to be as bad as it
looks at first sight.} Certainly, the result (\ref{garmal}) of
ref.\cite{malcolm} in the ALF limit is on a better standing after
having recovered it from a counterterm calculation.

Now we return to the result (\ref{nutact}), and specialize to nut
solutions using eqn.(\ref{nutvalues}):
\begin{equation}
I_{\rm TN-AdS}={4\pi n^2\over G}\left(1-{2 n^2\over l^2}\right)\ .
\end{equation}
For $n{=}l/2$ we recover the value for AdS$_4$ with boundary $S^3$,
which will be obtained and discussed in section~\ref{sec:revisit},
whereas for $n{=}0$ we recover the value (zero) for AdS$_4$ with
boundary $S^1{\times}S^2$. Again, these special cases may be regarded
as consistency checks on the internal consistency of our
implementation of the procedure.

Notice that the action becomes negative for
$n{>}n_0{=}l/\sqrt{2}$. More interestingly, being able to vary the
value of the Euclidean period $\beta{=}8\pi n$ we can compute the
energy of the solutions,
\begin{equation}
E=\partial_\beta I= {m_n\over G}\ ,
\end{equation}
which confirms the interpretation of $m$ as a mass parameter. We may
go further and compute the entropy and specific heat:
\begin{equation}\label{nutent}
S=\partial_\beta I -I= {4\pi n^2\over G}\left(1-{6 n^2\over l^2}\right)\
,
\end{equation}
\begin{equation}
C=-\beta\partial_\beta S= {8\pi n^2\over G}\left(-1+12{n^2\over
l^2}\right)\ .
\end{equation}
As had been already noticed in ref.\cite{cejm}, the mass (energy)
becomes negative for $n{>}l/2$. More strikingly, the entropy becomes
negative for $n{>}l/\sqrt{6}$ ! In particular, the entropy of AdS$_4$
($n{=}l/2$) is negative (equal to minus its action, since it has
$E{=}0$). Whereas a negative mass may not be too troublesome (one may
shift the energy scale), a negative entropy certainly would appear to
be a sign of pathological behaviour.  One should keep in mind, however, 
that this negative entropy appears because of a particular choice of (Euclidean) 
time coordinate. Even if it may seem  surprising at first sight that AdS$_4$ 
suffers from this pathology, we 
stress that this is a consequence of the particular choice of time slicing 
that we have made here, rather than an instrinsic property
of the AdS$_4$ solution itself.

In ref.\cite{gibbhawk} it was pointed out that in spaces where
Euclidean time is non--trivially fibered there appeared a contribution
to the entropy other than the usual one coming from the bolts (the
latter yields the black hole area law).  This extra entropy can be
associated to ``Misner strings''\cite{misner} (a geometric analog of
Dirac strings), and we would expect it to contribute to the entropy of
TN--AdS as $S_{\rm MS}{=}{A_{\rm MS}/(4G)}{-}\beta H_{\rm MS}$,  
\cite{hawkhunt1} where $A_{\rm MS}$ is the area of the string and $H_{\rm 
MS}$ is the Hamiltonian on it. Indeed, in the absence of a bolt this appears 
to be the only possible source of gravitational entropy for the TN--AdS 
solution. A brief calculation confirms that $S_{\rm MS}$ corresponds 
precisely to the expression we obtained in eqn.~(\ref{nutent}).

The fact that the specific heat becomes negative for $n{<}l/\sqrt{12}$
is an indication that the solutions become thermally unstable, making
them unusable for equilibrium thermodynamics\footnote{Nevertheless, a
negative specific heat is not so bad as a negative entropy; as a
matter of fact, as is well known, the Schwarzschild black hole in
asymptotically flat spacetime has negative specific heat---and so does the
ALF Taub--NUT solution.} (in the canonical ensemble). So if we declare
that the physically relevant solutions are those with both positive
entropy and positive specific heat, then the valid range for the nut
charge is
\begin{equation}
{l\over \sqrt{12}}\leq n\leq {l\over\sqrt{6}}\ .
\end{equation}
Solutions in this range have positive action and positive energy.

Finally, we note that the results for the energy, entropy and specific
heat of TB--AdS can be recovered by combining those for TN--AdS above,
and those for TB--AdS with the TN--AdS subtraction in ref.\cite{cejm}.

\subsection{Remarks upon Field Theory on Squashed Three--Spheres}

As discussed in ref.\cite{cejm,hawkhunt}, the study of solutions with
nut charge which are locally asymptotically AdS is relevant to the
$2{+}1$ dimensional ``exotic''\cite{exotic1} conformal field theories
which live on the world volume of M2--branes (and closely related
theories\footnote{Recall there is a problem with the spin structure of
TB--AdS, and so the M--theory interpretation is unclear
\cite{cejm}, although there is almost certainly a dual CFT
nonetheless.}),
after placing them on squashed three--spheres. Following that work, in
ref.\cite{dowker} the effective actions of various fields on squashed
three--spheres have been computed.

We do not expect to see in those particular field theory results any
signal of the apparently pathological behaviour ({\it e.g.}, negative
entropy) which we have found, and indeed we do not.  The difficulty
essentially lies in the fact that the field theory results can only be
used at weak coupling, whereas supergravity is describing a strongly
coupled regime of the field theory. The unusual behaviour belongs only
to the low temperature phase of the field theory, and strong coupling
effects change the picture drastically. Recall the phase structure
described in ref.\cite{cejm}:

$\bullet$ At high $T$ (small $n$) we have both TN--AdS and TB--AdS as
possible solutions, but the latter has the lower free energy, and is
therefore preferred. It was shown in ref.\cite{cejm} that at high $T$,
TB--AdS gives the expected behaviour $F \sim T^3$ which, not
surprisingly, is the result found in ref.\cite{dowker}. This is a
deconfined phase.

$\bullet$ At low $T$ (large $n$), however, the only existing phase is
TN--AdS. There is a phase transition separating this regime from the
deconfined phase mentioned above. This phase transition prevents us
from obtaining information from the results in ref.\cite{dowker},
since at weak coupling, where those results were obtained, one does
not get the confined phases.

\noindent
It is in this large $n$ region that the entropy becomes negative. In
fact, all of the negative entropy regime is within the region where
the only regular solution is TN--AdS: TB--AdS is absent there. 
One might speculate whether the Lorentzian version of the field theory 
(in this confined phase) contains ghosts that do not decouple. 
Such ghosts would yield a negative contribution 
to the entropy.

So we discover that the supergravity studies give us new information
on the strongly coupled phases of the theory on the world--volume of
the M2--brane, and related theories, after compactification on
squashed three--spheres.

\subsection{Flat and Hyperbolic Taub--NUT--AdS}
A solution where the nuts and bolts are flat planes instead of spheres
can
be found as well, and was analyzed in ref.\cite{cejm},
\be\label{flatnut}
ds^2 = V(r) \left(d\tau + {n\over l^2} (xdy-ydx)\right)^2 +
 {dr^2\over V(r)} + {r^2 -n^2 \over l^2}(dx^2 +dy^2)\ ,
\ee
where now,
\be
V = { - 2 mr  + l^{-2} (r^4 - 6 n^2 r^2-3
n^4)\over r^2 - n^2}\ .
\ee
The fibration is in this case a trivial one, and as a result the
Euclidean period $\beta$ is independent of $n$. Zero dimensional fixed
point sets of $\partial_\tau$ (``nuts'') exist for $m_{n}{=}{-}{4n^3
\over l^2}$. Solutions with bolts have a higher value of $m$. The
result for the counterterm calculation of the action for a solution
with generic $m$ and $n$ is
\be\label{flatnutac}
I={\beta L^2\over 8\pi G l^2}\left(ml^2-r_+^3+3n^2 r_+\right)\ ,
\ee
where, as usual, $r_+$ is the radial position of the fixed point set
($r_+{=}n$ for a nut), and $L^2$ accounts for the area of the $(x,y)$
plane, $-L/2{\leq}\{x, y\}{\leq}L/2$. It can be easily checked that
the action of ref.\cite{cejm}, where the nut solution was taken as a
reference background, can be recovered from (\ref{flatnutac}) as
$I({\rm bolt})-I({\rm nut})$. Moreover, (\ref{flatnutac}) is the same
result we would obtain had we performed a background subtraction
calculation with ``imperfect matching'' to AdS$_4$ (the latter in its
flat incarnation as $n{=}m{=}0$ in eqn.(\ref{flatnut})). We note that
for the nut values the action is negative, which reflects the fact
that its energy is negative---its entropy vanishes, as could have been
expected in the absence of bolts or Misner strings, so in fact we find
$I_{\rm nut}{=}\beta E_{\rm nut}$.

The last possibility is that of having hyperbolic fixed point sets of
$\partial_\tau$. The explicit solution is \be ds^2 = V(r) (d\tau + 2n
(\cosh\theta-1) d\varphi)^2 + {dr^2\over V(r)} + (r^2 -n^2)(d\theta^2
+\sinh^2\theta d\varphi^2)\ , \ee with \be V = { -(r^2 +n^2)- 2 mr
+l^{-2} (r^4 - 6 n^2 r^2-3 n^4)\over r^2 - n^2}\ .  \ee The fibration
is trivial, and again, there are no Misner strings.  However, it was
found in ref.\cite{cejm} that there are no hyperbolic nuts: {\it
i.e.,} it is not possible to make $r{=}n$ into a regular fixed point
of $\partial_\tau$. Nevertheless, bolt solutions can be
constructed. This is rather analogous to the situation we encounter
for hyperbolic black holes in section~\ref{sec:hyper}. The result for the
action is again formally very similar to (\ref{nutact}) and
(\ref{flatnutac}), \be I={\beta\sigma\over 8\pi Gl^2}\left( l^2 m+3
n^2 r_+-r_+^3\right)\ , \ee where $\sigma$ is the area of the
hyperbolic space (if quotients of $H^2$ are taken to yield surfaces of
genus $g{>}1$ (this is not essential) then $\sigma{=}4\pi(g{-}1)$).

\section{AdS Revisited}\label{sec:revisit}

Many of the quantities we have been computing can be translated into
field
theory results by using the dictionary provided by the AdS/CFT
correspondence
\cite{juan,brhe}, namely,
\be
c={3l\over2G} \quad {\rm for\; AdS}_3\ ,\quad
N^{3/2} \approx {l^2\over G} \quad {\rm for\; AdS}_4\ ,
\quad
N^2={\pi l^3\over 2 G}\quad {\rm for\; AdS}_5,\quad
N^3 \approx {l^5\over G} \quad {\rm for\; AdS}_7\ ,
\label{freedom}
\ee
where $c$ is the central charge of the dual CFT in two dimensions. The
powers of $N$ displayed above are measures of the number of
``unconfined'' degrees of freedom: for AdS$_5$, $N$ is the rank of the
gauge group of the dual ${\cal N}{=}4$ supersymmetric four dimensional
$SU(N)$ Yang--Mills theory. Meanwhile, for AdS$_4$ and AdS$_7$, the
dual field theories are the ones\cite{exotic2,exotic1} that describe
the world--volume dynamics of $N$ parallel M2--branes, and M5--branes,
respectively. The details of these latter two theories are still
rather indirectly and poorly understood, and the precise numerical
relationship between factors (missing in eqn.\reef{freedom} for these
cases) will not be needed here, as we will make no precise numerical
comparison.  While there is almost certainly a dual conformal field
theory for the case of AdS$_6$, we will not comment upon it
further. Note again that AdS for all of the dimensions listed are cases
that can be handled with the counterterms that we now have.

In section III, we considered the counterterm action for AdS with the
boundary geometries $S^1{\times}M_k^{n-1}$. In those cases, the action
is finite and interestingly for even $n$, an extra contribution
appears of the form $\beta E^0_k$ where $E^0_k$ is a constant energy
--- see eqn.~\reef{zero} in section IV. This constant energy is
readily interpreted in the dual field theory as a Casimir energy of
the conformal field theory on $S^1{\times}M_k^{n-1}$ --- see
ref.~\cite{positive} for another discussion of Casimir energies in the
AdS/CFT correspondence.  We can consider these results for $n{=}2,4$
in more detail: the well known Casimir energy of $(1{+}1)$ dimensional
CFT when going from the infinite plane to the cylinder
$\IR{\times}S^1$ is reproduced by the term $n{=}2$ in
eqn.~(\ref{zero}). Similarly, the Casimir energy of four dimensional
Yang--Mills theory on $\IR{\times}S^3$ is precisely the value of
$E_{k=+1}^0$ for $n{=}4$ \cite{pervijay}.

We found there as well that for the theory on $\IR{\times}H^3$, even
if $E_{k=-1}^0{\neq}0$, the result is consistent with the absence of a
Casimir energy after identifying correctly the ground state of the
theory. We remarked as well upon the striking appearance of a
non--zero entropy for this ground state, which strongly suggests the
presence of degrees of freedom which can contribute to the entropy but
not to the energy density, just like the ``precursor" states
identified in ref.~\cite{joelenny}. (This also reminds us of the
non--zero entropy extremal ground state studied in ref.\cite{cejm2}.)

We can translate some of our results for the cases of AdS$_4$ and
AdS$_7$ as well, finding that the Casimir energies derived by using
eqn.\reef{freedom} are correctly proportional to the number of degrees
of freedom in the theory, as can be deduced from the power of $N$
which appears in each case: The scaling with $N$ is precisely the same
as had been obtained from computations of black brane entropies
\cite{kletse}.

Let us now consider AdS$_{n+1}$ with boundary geometries $S^n$ and
$H^n$ as described by the metrics in eqn.~\reef{metric2}. In order to
notationally distinguish them from the family $S^1{\times}M_k$, we will
denote them with a ${}^\bullet_k$.  The results for the action are
somewhat more complicated to express for generic $n$ in an explicit
form.  For the three contributions (the bulk term, the
Gibbons--Hawking surface term, and the counterterm action) we find
\be
I^{\bullet}_{\rm bulk}={n\sigma_{k,n}\over 8\pi G l}\int_{r_+}^{r}
d\bar r {\bar
r^{n}
\over \sqrt{\bar r^2+kl^2}}\ ,
\ee
which can be expressed in terms of hypergeometric functions, but we
will only need its expansion for large $r$. The lower integration
limit is $r_+{=}0$ for $k{=}{+}1,0$, and $r_+{=}l$ for $k{=}{-}1$.
\be
I^{\bullet}_{\rm surf}=-{n\sigma_{k,n}\over 8\pi G l}r^{n}\sqrt{
1+k{l^2\over r^2}}\ ,
\ee
\be
I^{\bullet1}_{\rm ct}={(n-1)\sigma_{k,n}\over 8\pi G
l}r^{n}\ ,
\qquad
I^{\bullet2}_{\rm ct}={(n-1)\sigma_{k,n}\over 8\pi G
l}r^{n}\left({n\over 2(n-2)}{k l^2\over r^2}\right)\ ,
\qquad
I^{\bullet3}_{\rm ct}={(n-1)\sigma_{k,n}\over 8\pi G
l}r^{n}\left(-{n\over 8(n-4)}{k^2
l^4\over r^4}\right)\ ,
\ee
where we have separated the contributions of the individual terms
in the counterterm action \reef{explete}, as was done
in eqn.~\reef{contire}.
Again the limit $r{\rightarrow}\infty$ remains to be taken. Our
counterterms allow us to deal with $n{=}2,\ldots,6$. We therefore find
for $I^{\bullet}_{k,n+1}$
\be
I^{\bullet}_{k,3}=-k{l\sigma_{k,2}\over 16\pi G}\left(1+2\log{2r\over
l}\right)\ ,
\qquad
I^{\bullet}_{k,5}=k^2{3 l^3\sigma_{k,4}\over 64\pi G}\left( -1 +4\log
{2r\over l}
\right)\ ,
\qquad
I^{\bullet}_{k,7}=k{5l^5\sigma_{k,6} \over 64\pi G}\left( {5\over 4}
-3\log
{2r\over
l}
\right)\ ,
\label{evennn}
\ee
\be
I^{\bullet}_{k,4}={l^2 \sigma_{k,3} \over 4\pi G}\delta_{k,+1}\ ,
\qquad
I^{\bullet}_{k,6}=0\ ,
\ee
where we have omitted contributions which vanish in the limit
$r{\rightarrow}
\infty$. Here the most striking result is that for
even~$n$,~\reef{evennn} there remain logarithmically divergent
contributions from the bulk terms that are not cancelled by the
boundary counterterms. Furthermore given their logarithmic nature,
there is no way that they can be cancelled by a counterterm which is a
local integral over the boundary of a 
(polynomial) curvature invariant.  The
appearance of these divergences then presents a limitation for the
utility of the counterterm subtraction technique for investigations of
asymptotically AdS solutions in odd dimensions.\footnote{One could consider
the addition of nonpolynomial counterterms to resolve this problem.
A suitable counterterm would have the form $a_{n/2}({\cal R})
\log f({\cal R})$ where $a_{n/2}({\cal R})$ is the conformal anomaly
term (see below) and $f({\cal R})$ is an arbitrary curvature scalar.
While such a counterterm would render the action finite, 
it may produce problematic
results in calculating the boundary stress energy\cite{pervijay,bryk}.
We would like to thank Sergey Solodukhin for this suggestion.}

However, these divergences do not signal a problem for the AdS/CFT
correspondence, but rather provide a remarkable consistency check. The
possible existence of logarithmic divergences for odd spacetime
dimensions was noted in refs.~\cite{edadsii,sken}, where the
coefficients of the divergent terms were related to the conformal
anomaly in the dual field theory. It is a standard result of field
theory in curved spacetime\cite{birdav,conform} that the appearance of
a conformal anomaly in a classically conformally invariant theory is
due to logarithmic UV divergences (at least at the one--loop level)
appearing in the quantum field theory. Thus we have the UV/IR
relation\cite{UVIR} of the AdS/CFT correspondence at work here: the
appearance of an infinite volume singularity in the AdS calculation is
a reflection of the existence of a UV divergence in the CFT.

Further, if we make the association of the AdS radius with an energy
scale, we see that the divergence is logarithmic as required by the
field theory. For $n{=}4$, it is straightforward to verify that in fact
the ${\cal N}{=}4$ SYM theory has a conformal anomaly on $S^4$ or
$H^4$, and further a perturbative weak coupling calculation reveals a
logarithmic singularity in the effective action for the background
metric at one--loop\cite{conform,private}. That is, despite the
remarkable finiteness properties of ${\cal N}{=}4$ SYM to higher loops
in flat space\cite{loop}, in curved spacetimes the ${\cal N}{=}4$
supersymmetry is only enough to protect against potential quadratic
and linear divergences. In general though, there is the possibility of
one--loop logarithmic divergences.  One can show though that for the
${\cal N}{=}4$ SYM theory, the coefficient of these divergent terms
will always vanish on product space geometries\cite{private}. This is
consistent with the fact that no logarithmic singularities were
found in the actions \reef{even} for the boundary geometries
$S^1{\times}M_k$.

Let us make this connection somewhat more precise. In the presence of a 
trace anomaly term $T^c_c$ the action picks a divergent contribution 
of the form
\be\label{logterm}
I_{\rm log}=\left(\log{r\over l}\right)\int d^n x\sqrt{h} T^c_c
\ee
(see, {\it e.g.}, \cite{sken}. The cutoff $\epsilon$ in that paper is related 
to ours as $\epsilon{=}(l/r)^2$). Therefore we would expect, and we will 
actually verify it below, that the logarithmic terms we have found follow 
directly from the value of the anomaly. 

Let us now write some of the results (\ref{evennn}) in terms of field theory
parameters, in order to make a comparison with the field--theoretical 
expression (\ref{logterm}).  The result for $n{=}4$ should be related to 
${\cal N}{=}4$ supersymmetric four--dimensional Yang--Mills theory on
de Sitter ($S^4$) or anti--de Sitter spacetime ($H^4$). We find a
pleasingly simple result,
\be\label{Ik15}
I^{\bullet}_{k,5}=-k^2{N^2\over 4}\left( 1 -4\log {2r\over l} \right)\ .
\ee
Note the fact that the action does not change sign when going from
$S^4$ to $H^4$ ({\it i.e.,} $k{=}{+}1\rightarrow k{=}{-}1$) has its
counterpart in the field theory in the fact that the divergent term in
the effective action is given by curvature squared terms. In fact,
this result generalizes to no sign change for $n{=}4p$ where the
conformal anomaly is proportional to the $2p$ power of curvatures, and
a change of sign for $n{=}4p{+}2$, where the relevant power is
$2p{+}1$ \cite{conform}.

Explicitly, for ${\cal N}{=}4$ SYM on $S^4$ the trace anomaly 
is \cite{birdav}
\be
T_c^c={\sum_s q(s)\over 240\pi^2 l^4}={3 N^2\over 8\pi^2 l^4}
\ee
where $q(s)$ measures the contribution of spin $s$ fields: for ${\cal N}{=}4$ 
SYM the result, for large $N$, is $\sum_s q(s){=}90 N^2$. Plugging 
this expression in (\ref{logterm}) we recover the exact logarithmic term in 
(\ref{Ik15}). 

The finite part of the action would be expected to follow from 
field--theoretical calculations as well. The scaling~$N^2$ is just expected 
from the number of degrees of freedom of the theory, and the
absence of any other factors follows from dimensional
arguments. Related to this is the fact that the trace anomaly
$<\!T^c_c\!>$ can be computed exactly within the AdS/CFT
correspondence\cite{sken}. Having that, the full stress tensor follows
in this case since the symmetry of the geometry will dictate that
$<\!T_{ab}\!>= h_{ab}<\!T^c_c\!>/n$. Therefore, it is not surprising
that a calculation of the stress tensor in the manner described in
ref.~\cite{pervijay} reproduces this result.

For AdS$_3$ we can write the result as \be I^{\bullet}_{k,3}=-k{c\over
6}|g{-}1|\left(1+2\log{2r\over l}\right)\ .  \ee where $g$ is the
genus of the two--dimensional boundary surface, {\it i.e.}, for the
hyperbolic case we have taken quotients by discrete groups in order to
find genus $g$ surfaces (this is not essential).  Again, the
logarithmic term is precisely the result for a $(1{+}1)$ dimensional
conformal field theory on a surface of genus $g$, area $4\pi l^2
|g{-}1|$, as follows from the trace anomaly on such a surface,
$T^b_b{=}{-}kc/(12 \pi l^2)$.

In the same vein, we would expect that the presence of a
logarithmically divergent factor for AdS$_7$ can be interpreted in
terms of the effective field theory for the M5--brane when defined on
six dimensional de Sitter space. The anomaly for this theory has not
been computed by independent field--theory methods, rather it has been
deduced in ref.~\cite{sken} using the AdS/CFT correspondence. Using
that result, the logarithmic term comes out precisely as expected.

It is clear that in the present paper we have only scratched the
surface of the full subject, and more detailed and extensive
comparisons between the results of Euclidean quantum gravity and the
dual field theories are possible. We hope to report on progress on
this in the future.

\section*{Acknowledgments}
RE is supported by EPSRC through grant GR/L38158 (UK), and by grant
UPV 063.310--EB187/98 (Spain).  Support for CVJ's research was
provided by an NSF Career grant, \#PHY9733173.  RCM's research was
supported by NSERC (Canada) and Fonds FCAR du Qu\'ebec.  This paper is
report \#'s EHU--FT/9906,
DTP--99/21, UK/99--04 and McGill/99--12.  We would like to thank Vijay
Balasubramanian, Cliff Burgess, Andrew Chamblin, Ivo Sachs and Joe
Straley for comments and useful conversations.

\end{document}